\documentclass[aps,twocolumn]{revtex4}
\usepackage{amssymb}
\usepackage{amsmath}
\usepackage{epsfig}
\begin{document}

\title{Energy transmission in the forbidden bandgap of a nonlinear chain}
\author{F. Geniet, J. Leon}
\affiliation{Physique Math\'ematique et Th\'eorique, CNRS-UMR5825\\
34095 MONTPELLIER Cedex 05 (FRANCE)}

\begin{abstract} A nonlinear chain driven by one end may propagate energy in
the forbidden band gap by means of nonlinear modes. For harmonic driving at a
given frequency, the process ocurs at a threshold amplitude by sudden large
energy flow, that we call {\em nonlinear supratransmission}.  The bifurcation
of energy transmission is  demonstrated numerically and experimentally on the
chain of coupled pendula (sine-Gordon and nonlinear Klein-Gordon equations) and
sustained by an extremely simple theory.\end{abstract}

\maketitle

\paragraph*{Introduction.}

A nonlinear chain of oscillators has a few striking fundamental properties,
the first of which, known as the Fermi-Pasta-Ulam recurence phenomenon
\cite{fpu} is the {\em spectral localization} of energy: injected initially  in
one given eigenmode energy does not eventually distributes amongst higher modes
(as one would normally expect from a quasi linear approach).  The FPU discovery
has been at the origin of nonlinear studies all over the world and led to the
birth of the {\em soliton} concept \cite{zabusky}, soon followed by the
creation of the {\em inverse spectral tranform} (IST) and the concept of {\em
integrability} \cite{kruskal} \cite{zakshab} \cite{akns}.

A nonlinear chain also shows up {\em spatial localization} of energy in the
form of nonlinear coherent structures, the solitons \cite{solitons}. This
universal behavior is well understood in the concept of integrability: any
localized bounded initial condition eventually evolves to a number of isolated
solitons and a vanishing background of quasi-linear radiation.

Nonlinear energy localization is now deeply studied in the context of discrete
systems where {\em intrinsic nonlinear modes} (breathers for short) have been
shown to be the fundamental basic objects \cite{discrete,aubry}, and have been
experimentally observed in {\em Josephson ladders} \cite{ladder}.  Moreover, as
an effect of non-integrability, nonlinear modes do exchange energy and the
spatial localization acquires novel quite interesting features \cite{peyrard}.

The behavior of a nonlinear medium submitted to boundary data, as opposed to
initial conditions, is also of fundamental interest. One famous instance is the
project of data transmission in optical fibers in nonlinear Kerr regime for
which the basic model is the nonlinear Schr\"odinger equation \cite{has-moll}. 
Another instance is the nonlinear property of self-induced transparency (SIT)
of a two-level system submitted to high-energy incident (resonant) laser pulse
\cite{machan} \cite{akn}.  In those two cases the concept of integrability has
proved its efficiency as the physical boundary value problem maps to a well
posed Cauchy (initial value) problem.

There, the nonlinear coherent structures emerge from  sufficiently
energetic localized input pulse. Another fundamental question is then the
behavior of a medium submitted to continuous wave radiation.  Such problems
have been studied so far mainly through peturbation of integrable models by
{\em external driving}, see e.g. \cite{kivshar} for a review and
\cite{ext-dr} for interesting recent
developpments. But the more basic problem of scattering of continuous wave onto
nonlinear media has not been much studied.  Preliminary numerical results have
been obtained recently in \cite{jg-alex}, where a linear monochromatic wave is
scattered on a nonlinear medium. The nonlinearity allows then wave transmission
under nonlinear modes generation, with a threshold which was not given a
theoretical ground.  

\begin{figure}[ht]
\centerline{\epsfig{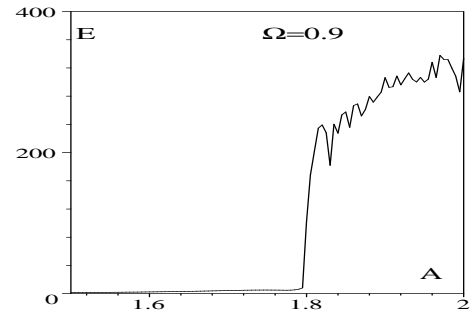}}
\caption{Plot of the energy $E$ injected in the sine-Gordon chain for $T=140$ 
as a function of the driving amplitude $A$.}
\label{fig:energy09}\end{figure}

Those problems share a common basic question: the response of a nonlinear
medium to periodic boundary data.  We demonstrate here the existence of a
bifurcation of wave transmission within a forbidden band gap (FBG) in a
nonlinear chain forced (periodically) at one end.  The related brutal
energy flow through the medium is illustrated on figure \ref{fig:energy09}
where the energy $E$ penetrating the medium is plotted as a function
of the amplitude $A$ of the boundary driving at a frequency in the FBG
(see later for details).

The mechanism of this bifurcation takes its origin in nonlinear modes
generation by the periodic forcing and allows for energy injected at one end of
a chain to penetrate a medium by an intrinsic nonlinear process.  This is
called {\em nonlinear supratransmission} and it is illustrated  both by
numerical simulations of sine-Gordon and of nonlinear Klein-Gordon equations,
and by experiments on a mechanical chain of pendula.

We shall first prove that, for sine-Gordon, the mechanism for nonlinear modes
generation follows a very simple rule providing an explicit formula for the
bifurcation diagram in the parameter space $\{A,\ \Omega\}$ where $A$ is the
amplitude of the driving and $\Omega$ its frequency. The rule states that
energy penetrates the medium as soon as the amplitude $A$ of the harmonic
driving at frequency $\Omega$ exceeds the maximum amplitude of the static
breather of same frequency. Then the system adapts to the driving boundary by
means e.g. of a 2-breather solution whose second component cannot be {\em also}
static (a 2-breather solution of sine-Gordon is made of 2 breathers of {\em
different velocities}) and hence propagates in the medium. The whole process
eventually repeats itself.

Nonlinear supratransmission holds also for the nonlinear Klein-Gordon chain
obtained by Taylor expansion of sine-Gordon. Then this process does not relie
on integrability and is expected to be generic as soon as the model possess a
natural forbidden band gap.

We discover then that {\em harmonic phonon quenching} enhances nonlinear
penetration : when the first significant harmonic generated by the nonlinearity
(the third one in sine-Gordon) falls inside the FBG, the related phonons stick
on the boundary and  contribute to the driving allowing supratransmission at
lower amplitude.  This is an interesting property in view of experimental
applications.

Last the energy transmission by means of nonlinear modes generation is explored
and the bifurcation is described  by expressing the energy injected in the
medium in terms of the driving amplitude, hence furnishing a striking view of
the supratransmission process.

\paragraph*{Model.}

Consider the discrete sine-Gordon chain of coupled oscillators $u_n(t)$
(time is normalized to the eigenfrequency of the individual oscillator)
\begin{equation}\label{SG}
\ddot u_n-c^2(u_{n+1}-2u_n+u_{n-1})+\sin u_n=0,\end{equation}
on a semi-infinite line $n>0$ with given initial-boundary value problem,
namely the data of the driving boundary $u_0(t)$, the initial positions
$u_n(0)$, initial velocities $\dot u_n(0)$ and boundary condition at the
chain end. The linear dispersion relation $\omega(k)$ is given by
\begin{equation}\label{disp}
\omega^2 =1+2c^2(1-\cos k).
\end{equation}
The chain will be submitted to external harmonic forcing $u_0(t)=A\sin\Omega t$
on a medium initially at rest.  For a frequency $\Omega$ in the phonon band,
quasi-linear waves are generated in the medium and, for large enough amplitude,
these waves will undergo Benjamin-Feir instability hence creating localized
excitations. These nonlinear modes have a very important role in the large time
asymptotic properties of a nonlinear system  and are suspected to be
responsible for turbulent-like behavior \cite{michel}.

We consider here a driving frequency in the FBG, namely $\Omega<1$, 
for which the linear theory would lead to the evanescent wave 
$A\sin(\Omega t)\exp[-\lambda n]$ with $\lambda$ given by 
\begin{equation}\label{lambda}
\lambda={\rm arccosh}\left(1+\frac{1-\Omega^2}{2c^2}\right).\end{equation}
In the nonlinear case, in order to fit the boundary condition \eqref{forcing},
the medium {\em adjusts a static breather}
\begin{equation}\label{stat-breath}
u_b(n,t)=4\arctan\left[\frac{\lambda c\sin(\Omega t)}
{\Omega\cosh(\lambda(n+n_0))}\right]
\end{equation}
which is an exact solution in the continuous limit only but works well for
strong coupling (the fully discrete case $c\ll 1$ will be considered
separately).

\paragraph*{Bifurcation process.}

Adjusting a static breather actually means to adjust the value of the breather
center $-n_0$ such that the oscillation amplitude at the boundary $n=0$ matches
the forcing amplitude.  This works up to the maximum value $A_s$ of the
breather amplitude realized for $n_0=0$. Hence from \eqref{stat-breath}, the
threshold $A_s$ reads as the following function of the frequency $\Omega$
\begin{equation}\label{threshold}
A_s=4\arctan\left[\frac{c}{\Omega}{\rm arccosh}\left(1+\frac{1-\Omega^2}{2c^2}
\right)\right],\end{equation}
which has the accurate simplified continuous approximation 
$4\arctan[\sqrt{1-\Omega^2}/\Omega]$.

This bifurcation threshold is now checked on numerical simulations of 
\eqref{SG} with the following initial-boundary conditions
\begin{equation}\label{forcing}
u_0(t)=A\sin\Omega t,\quad u_n(0)=0,\quad 
\dot u_n(0)=A\Omega e^{-\lambda n}.\end{equation}
The initial velocities are those of an evanescent wave such as to partly avoid
the shock wave generated by vanishing initial velocities (the same results, but
time consuming, are obtained for vanishing initial velocities and a driving
amplitude smoothly growing from the value $0$ to $A$). Finally, an infinite
medium is simulated by an absorbing boundary. The simulations are made with the
{\tt dsolve} routine in MAPLE with $10^5$ maximum iterations. The absorbing end
consists in adding a damping $\gamma \dot u_n$  in the model, with intensity
$\gamma$ slowly varying from 0 to 2 on the last 50 particules.

\paragraph*{Results.}
\begin{figure}[ht]
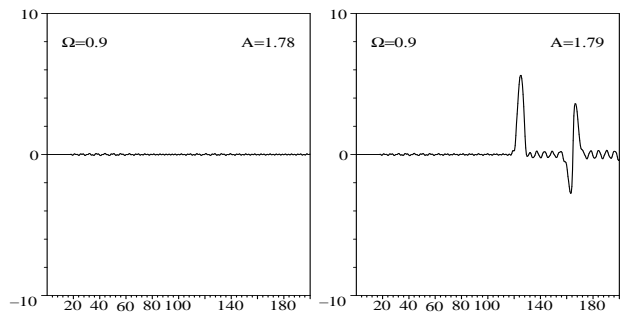

\centerline{\epsfig{file=noth.eps,height=4cm,width=4cm}
            \epsfig{file=breath.eps,height=4cm,width=4cm}}
\caption{Function $u_n(t)$ for $n=60$ in the case $\Omega=0.90$.
The amplitude are $A=1.78<A_s$ for the first figure and
$A=1.79=A_s$ for the second.}\label{fig:motions}\end{figure}

In order to generate a bifurcation diagram one has to compare between
simulations where the nonlinear supratransmission does or does not occur, as
illustrated on figures \ref{fig:motions} where the motion of one particule of
the chain is plotted for driving amplitudes just below and just above the
threshold.  The simulation is performed with $200$ particles with a coupling
$c=4$. Each large oscillation in the secong figure corresponds to a breather
(constituted of a single hump oscillating and propagating) passing by. Two of
them are generated and cross the site $60$ at times $120$ and$160$. The small
oscillation seen between the humps are the harmonic phonons, mainly of
frequency $3\Omega$.

\begin{figure}[ht]
\centerline{\epsfig{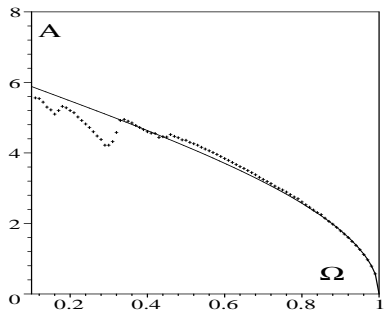}}
\caption{ Bifurcation diagramm in the $(A,\Omega)$ plane. The curve is the 
graph of formula \eqref{threshold}. The points indicate the lowest amplitude 
for which nonlinear supratransmission starts.}\label{fig:bif}\end{figure}

We may now proceed with a systematic exploration of the chain response. The
result is presented on figure \ref{fig:bif} obtained for $200$ particles with a
coupling $c=10$ (some experiments have been actually made with smaller coupling
and less number of points to shorten computation times) for a typical time of
$200$ (for frequencies close to the gap value $1$, time had to be increased up
to $500$). The points on figure \ref{fig:bif} are obtained with an absolute
precision of $10^{-2}$ for the amplitude $A$. They are compared to the
theoretical threshold expression \eqref{threshold} (continuous curve).

\paragraph*{Harmonic phonon quenching.}

The figure \ref{fig:bif} shows excellent agreement to formula \eqref{threshold}
at least in the region $0.34<\Omega<1$. Discrepencies are seen to occur
starting below $0.33$ and $0.18$. This results from the driving which, thanks
to the nonlinearity, generates phonons at multiple frequencies (here third and
fifth). If these frequencies lie in the phonon band, the phonons move away from
the boundary and have no effect on the forcing. If however they lie in the FBG,
the related phonons do not propagate  (which we call phonon quenching) and
stick on the boundary where they add contribution to the driving.

This effect should thus disappear if the phonons are eliminated by driving the
boundary with the exact breather expression \eqref{stat-breath} used to
calculate  $u_0(t)$, $u_n(0)$ and $\dot u_n(0)$. In that case we have checked
that nonlinear supratransmission {\em never occurs} at an amplitude $A<A_s$,
while it occurs  for very small deformation of the perfect breather. For
instance, by using $(1+\epsilon)u_b(0,t)$ as driving boundary, in the case
$\Omega=0.30$,  supratransmission occurs for $\epsilon =6\ 10^{-4}$, i.e.  for
a driving amplitude $A=5.0674$ instead of the threshold  $A_s=5.0644$.

\paragraph*{Energy transmission.}

The nonlinear supratransmission allows energy to flow through the medium and we
compare here this energy flow below and above the threshold. The energy
injected in the medium by the driving boundary  is given  at time $T$ by
\begin{equation}\label{energy}
E=-c^2\int_0^Tdt\ \dot u_0(t)[u_1(t)-u_0(t)]\ .\end{equation}
In our case $u_0(t)$ is the driving \eqref{forcing} and the chain is supposed
infinite with $u_n(t)\to0$ as $n\to\infty$.  Chosing for $T$ an integer
multiple of the period of excitation makes this energy to vanish identically in
the linear case if the driving frequency falls in the FBG.
  
In the nonlinear case, expression \eqref{energy} is computed  numerically.  For
a driving frequency $0.9$ and amplitudes running from $1.5$ to $2.0$, we obtain
the  figure \ref{fig:energy09} where the bifurcation is seen to occur for
$A=1.80$,  the value predicted by fromula \eqref{threshold}. This simulation
has been runned for frequencies in the range $[0.2,0.9]$, with expected results.

\paragraph*{Nonlinear Klein-Gordon.}
\begin{figure}[ht]
\centerline{\epsfig{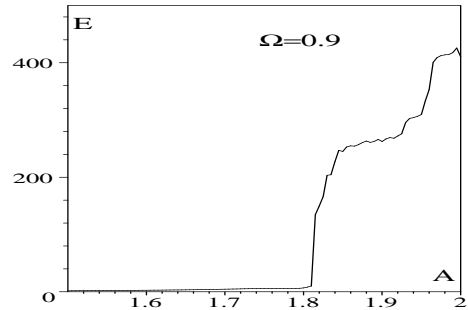}}
\caption{Energy $E$ injected in the Klein-Gordon nonlinear chain 
for $T=140$ as a function of the driving amplitude $A$.}
\label{fig:kg-09}\end{figure}
The approach stems from the existense of a breather solution of
the model equation, allowing to determine the threshold amplitude. Then a
fundamental question is the dependence of nonlinear supratransmission on some
{\em integrability} property of the equation. To give an indication that this
process is generic (with a stop gap), we have performed numerical simulations
of the following nonlinear Klein-Gordon chain:
\begin{equation}\label{KG}
\ddot u_n-c^2(u_{n+1}-2u_n+u_{n-1})+ u_n-\frac1{3!}u_n^3+
\frac1{5!}u_n^5=0,\end{equation}
the Taylor truncated expansion of sine-Gordon (the fifth order
is kept to ensure a confining potential at large $u_n$).

Then this system is solved with the boundary driving (\ref{forcing}) and the
energy (\ref{energy}) is computed for the same parameter values as for figure
\ref{fig:energy09}. The result is displayed on figure \ref{fig:kg-09}. 
Nonlinear supratransmission is seen to still occur, though a nonlinear mode
solution of the model does not exist. By scanning the frequency range in the
gap, we have obtain that the process occurs down to $\Omega=0.7$ and then
disappears.

\paragraph*{Experiment.}

\begin{figure}[ht]
\centerline{\epsfig{file=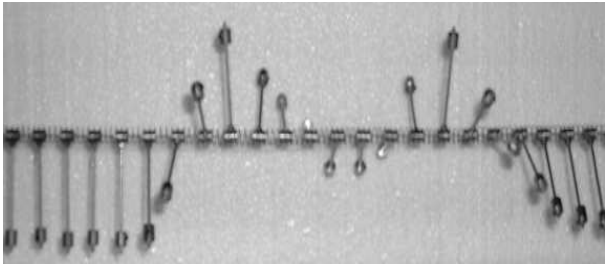,height=3.5cm,width=8cm}}
\caption{Picture of a breather generated in a mechanical pendula chain
driven at one end at a frequency in the forbidden band gap.}
\label{fig:pendule}\end{figure}

The phenomenon of nonlinear supratransmission can be experimentaly realized on
a mechanical pendula chain driven at one end by a periodic torque
\cite{alexandra}.  The detail analysis of such experiments  will be published
later, but it is worth showing here a picture of a breather generated by the
boundary driving at a frequency inside the FBG.  The breather on picture
\ref{fig:pendule} has been obtained with a chain of 30 pendula of angular
eigenfrequency $\omega_0=15\ Hz$ (upper value of the FBG) by driving at
(angular) frequency $12.7\ Hz$, which in the normalized units used here
corresponds to $\Omega=0.85$. The coupling constant has been measured to be
$c=32$.

\paragraph*{Conclusion.}

A novel fundamental property of a nonlinear medium has been unveiled, namely
the capacity to transmit energy under irradiation in a forbidden band gap by
means of nonlinear mode generation. Theoretical construction of the bifurcation
diagram is extremely simple when the one-breather solution is known and the
numerical simulations fit strikingly well the theory.

There remain of course many interesting open questions and currently under
study are the modulation of the driving signal, the effect of damping,
viscosity, disorder, external bias, discreteness, etc... Another essential
point is the search of nonlinear supratransmission in others physical
situations.  For instance this result might provide understanding of the very
mechanism of the  {\em generation} of gap solitons in photonic band gap
materials \cite{mills,sterke}.

Last but not least, as the sine-Gordon chain is a model for discrete Josephson
transmission lines \cite{ustinov}, we expect interesting applications in this
field. There, the boundary driving is realized by microwave irradiation through
a finline antenna \cite{ust-myg}, and it actually corresponds to prescribing
the {\em derivative} at the origin (Neuman condition). Preliminary numerical
simulations has shown that nonlinear supratransmission also works and that the
bifurcation process obeys the same type of simple rule. Detailed results on
this question wil be published later.

\paragraph*{Aknowledgements.} It is a pleasure to aknowledge enlightening
discussions with M.J. Ablowitz, S. Aubry, G. Kopidakis, M. Schuster and A.V.
Ustinov. This work has received support of contract INTAS 99-1782.

\end{document}